\documentclass[a4paper,onecolumn]{emulateapj}

\usepackage{booktabs}

\usepackage{latexsym}
\usepackage{enumerate}

\usepackage{amstext}
\usepackage{apjfonts}
\usepackage{amsmath}
\usepackage{amssymb}

\usepackage{footmisc}

\usepackage{natbib}
\usepackage{rotating}
\usepackage[colorlinks=true,linkcolor=blue,citecolor=blue]{hyperref}
\usepackage{changepage}
\usepackage{scrextend}
\usepackage[ampersand]{easylist}
\usepackage{xcolor}

\newcommand*{\cf}{\fontfamily{lmtt}\selectfont}

\begin{document}

 \title{Houdini for Astrophysical Visualization}

\author{J.P. Naiman\altaffilmark{1}, Kalina Borkiewicz\altaffilmark{2}, A.J. Christensen\altaffilmark{2}}
\altaffiltext{1}{Harvard-Smithsonian Center for Astrophysics, The Institute for Theory and Computation,
60 Garden Street, Cambridge, MA 02138}
\altaffiltext{2}{Advanced Visualization Laboratory, National Center for Supercomputing Applications, 1205 West Clark Street, Urbana, IL 61801}

 \begin{abstract}

 The rapid growth in scale and complexity of both computational and observational astrophysics over the past decade necessitates efficient and intuitive methods for examining and visualizing large datasets.  Here we discuss some newly developed tools to import and manipulate astrophysical data into the three dimensional visual effects software, {\it Houdini}.  This software is widely used by visual effects artists, but a recently implemented Python API now allows astronomers to  more easily use {\it Houdini} as a visualization tool.  This paper includes a description of features, work flow, and various example visualizations.  The project website, {\cf www.ytini.com}, contains {\it Houdini} tutorials and links to the Python script Bitbucket repository aimed at a scientific audience to simplify the process of importing and rendering astrophysical data.
 
\end{abstract}


\section{Introduction}

Astronomers have long used visualizations of their observed and simulated data to stimulate the public's interest in science.  Recent inroads which utilize three dimensional modeling and game development software make the possibility of outreach materials generated by a myriad of individual scientists, rather than a few graphics studios, an exciting new avenue to be explored \citep{kent2015,taylor2015,naiman2016}. As technological advancements in graphics and gaming progress, the scientist is presented with innovative methods to further develop their own public outreach  \citep{vogt2013,steffen2014,vogt2014,brown2014,madura2015}.

In this paper, we introduce several astrophysical data processing Python tools for use in {\it Houdini}, the three dimensional visual effects software used by a wide variety of professional studios.  Here, we make use of {\it yt} \citep{turk2011} as a data reader within {\it Houdini} to serve as a bridge between the astronomical and graphics community.  A variety of resources including tutorials, example scientific {\it Houdini} files and workflows, Python scripts, both raw and preprocessed volumetric scientific data sets, and links to external resources have been compiled into the newly launched ``Houdini For Astronomy" website, {\cf www.ytini.com}, as a resource for the astronomy community. Integration of astronomical data within {\it Houdini} provides scientists with the ability to create production quality visualizations that encompass not only images in papers and on websites, but large scale movies for dome shows and various virtual reality devices.

 The organization of this paper is as follows: after a brief overview of the {\it Houdini} software and its use in professional graphics studios in section \ref{section:houdini}, an introduction to the {\it Houdini} GUI and an example workflow including the import of astrophysical data and image rendering is presented in section \ref{section:workflow}.  The different methods for importing datasets are discussed in section \ref{section:inputFormats}.  
The paper concludes with example renders and interactive three dimensional models along with a brief discussion of their possible extensions with visual effects methods in section \ref{section:examples}.  We conclude with a summary and discussion of future plans in section \ref{section:discussion}.

\section{Houdini Background and Usage} \label{section:houdini}

{\it Houdini} is a commercially available visual effects software package that is widely used by visual effects studios in Hollywood, video game development studios, and other graphics industries. Many of these studios and companies, such as Disney/Pixar, Dreamworks, Double Negative, Method, Digital Domain, Framestore, Axis, Electronic Arts, and Ubisoft, contribute to a lively user and developer community, and in concert with exhaustive documentation by the developer, have made the software accessible to smaller research communities 
such as the research teams at the University of Technology Sydney, the California Academy of Sciences, the American Museum of Natural History, and the National Center for Supercomputing Applications.\footnote{University of Technology Sydney: http://www.uts.edu.au/, California Academy of Sciences: http://www.calacademy.org/, American Museum of Natural History: http://www.amnh.org/, National Center for Supercomputing Applications, Advance Visualization Laboratory: http://avl.ncsa.illinois.edu/}  
These research teams use the extensive visual effects methods available in {\it Houdini} to craft educational documentaries and planetarium shows from observed and simulated datasets from a wide variety of scientific disciplines. 

While several other three dimensional graphics packages exist to render astronomical data ({\it Maya}, {\it Blender}), {\it Houdini} provides some advantages for the rendering of multidimensional scientific datasets.  In addition to being an industry standard software, increasing the ease of collaborating with professional filmmakers, the native renderer in {\it Houdini}, Mantra, is considered to be one of the highest-quality renderers for volumetric data, which is a data format often generated by both computational and observational astronomers.  Packages such as {\it Maya} and {\it Blender} provide excellent general purpose tools for three dimensional modeling, compositing, and animating.  However, its {\it Houdini}'s focus on rendering specifically for visual effects which allows for the production of visually compelling volume renders as the multidimensional nature of astrophysical data makes it more closely akin to industry standard visual effects frameworks then either modeling or animating frameworks.  In addition, {\it Houdini} offers many methods for arbitrary data types that most animation software does not, including pre-built nodes for importing data from tables and easy conversions between imported data types to compare visualizations from differently formatted datasets.

{\it Houdini} is unique among commercial visual effects tools as its scene development paradigm is intended to be modular, procedural, and easily altered at any stage of development. This proceduralism is represented as a network of operator nodes that affect geometry, shaders, rendering, etc. The node network should feel familiar to data scientists, as it is a graphical representation of the flow of information through blocks of code. In addition to a programmer-friendly operator network, there are nodes that allow scripting with a native optimized C-like language called VEX, there are nodes that allow Python scripting, and there are many other ways to extend the native capabilities of the software including external so-called ``Digital assets" and the C++ Houdini Development Kit.  Developers familiar with software such as {\it Maya} or {\it Blender} will be familiar with the way this concept is implemented by those programs in their shader network interfaces, however, for the novice we review several vital user interfaces in section \ref{section:workflow}.

All the widely used operating systems are supported by {\it Houdini}, including Windows, Linux, and Mac. Although there is a high-priced license available to studios, significantly discounted licenses are available to educational institutions, and a free learning edition with few limitations is available for direct download from the developer, SideFX\footnote{www.sidefx.com}. 

Scene layout and geometry data are stored within {\it Houdini} as a series of nested directories, and represented this way within the interface. At the scene level, one can enter a directory that contains all of the surface data, a directory for the render data, a directory for the shader data, etc. A user then builds operator networks within these directories. Surface directory networks are built from surface operators or SOPs, render networks are built from ROPs, shader networks are built from SHOPs, and so on. New nodes can be placed by pulling up menus or tool shelves. The Tab key menu is also an efficient way to peruse the available operators.

While scenes can become increasingly complicated for production-level renders, the rendering of an image of astronomical data can be accomplished in relatively few steps.

\section{Example Houdini Workflow} \label{section:workflow}

The pathway from an astrophysical dataset to a rendered image and/or three dimensional model revolves around first loading the data in a format that is understood by {\it Houdini},  placing the objects in the three dimensional space, determining how light will travel through the dataset based on user defined shaders, and then exporting a rendered image(s) or a three dimensional model.  

\begin{figure*}
\centering
\includegraphics[width=1.0\textwidth]{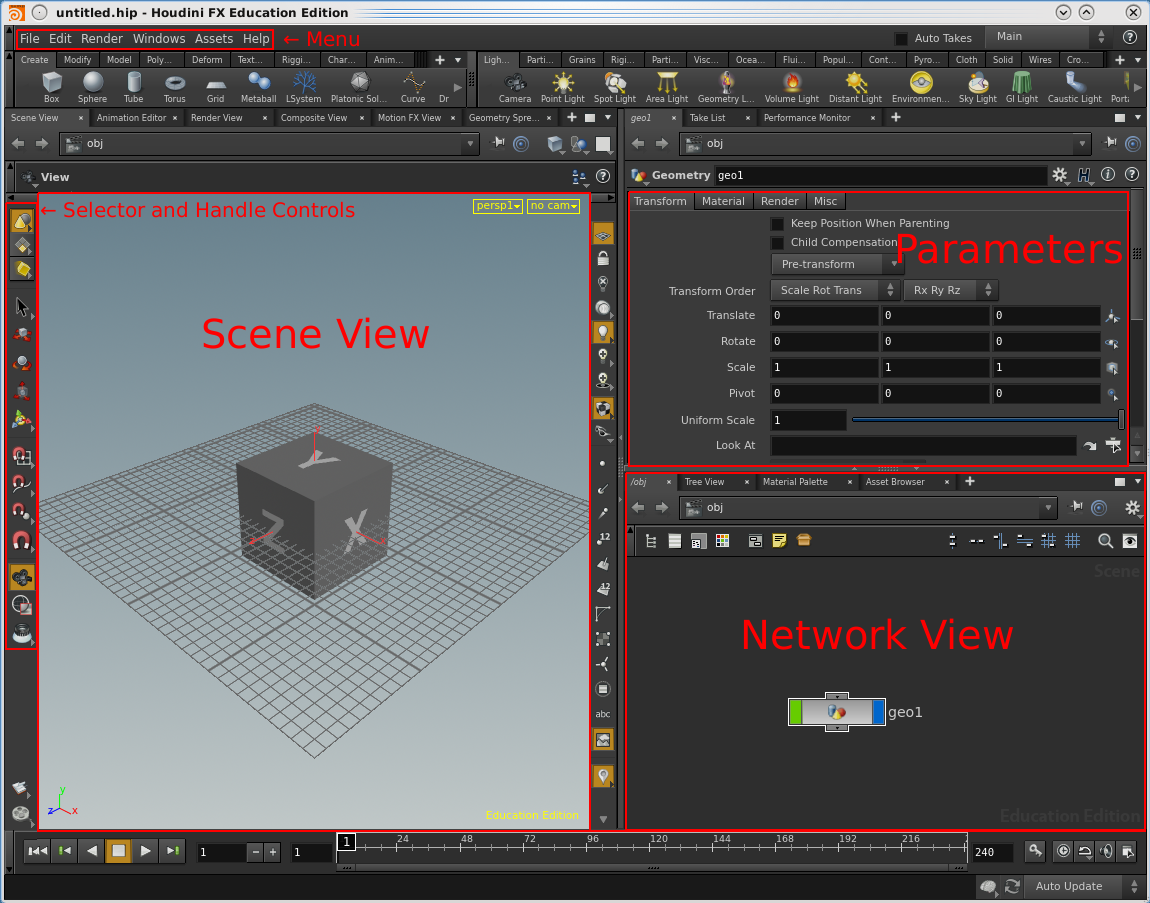}
\caption{A typical {\it Houdini} GUI includes a window for direct interaction with data and objects in the three dimensional space (Scene View), an interactive window for changing parameters associated with each three dimensional object (Parameters panel) and a visual representation of the interplay between three dimensional objects, shaders, and file systems (Network View).  Each panel has a variety of sub-panels.  Two of the most used panels, which allow for access to external files (Menu) and finer control of three dimensional objects in the Scene View (Selector and Handle Controls) are also highlighted.}
\label{fig:gui}
\end{figure*}

While the data handling portion of the workflow is done through the custom Python scripts discussed in section \ref{section:inputFormats}, the interaction with the loaded data and {\it Houdini} cameras is accomplished through the GUI, shown in Figure \ref{fig:gui}.  
The GUI can be modified to include further informational and analysis panels, but the default version shown in Figure \ref{fig:gui} highlights the key interfaces for the novice user - the Scene View provides direct three dimensional interaction with the loaded data, the Network View depicts the connected network of data and data modifiers and shaders which determine the final rendered image, and the Parameters panel provides information about each node in the network.  
The smaller highlighted sections include the Menu bar which is used to import code and data, the Selector and Handle Controls which allow the user different methods to select and move data within the three dimensional space of the Scene View.

Using interactions with this basic GUI we outline the general workflow of a {\it Houdini} session, but caution that this workflow is highly variable between users and projects.

\begin{figure*}
\includegraphics[width=1.3\textwidth, angle=90]{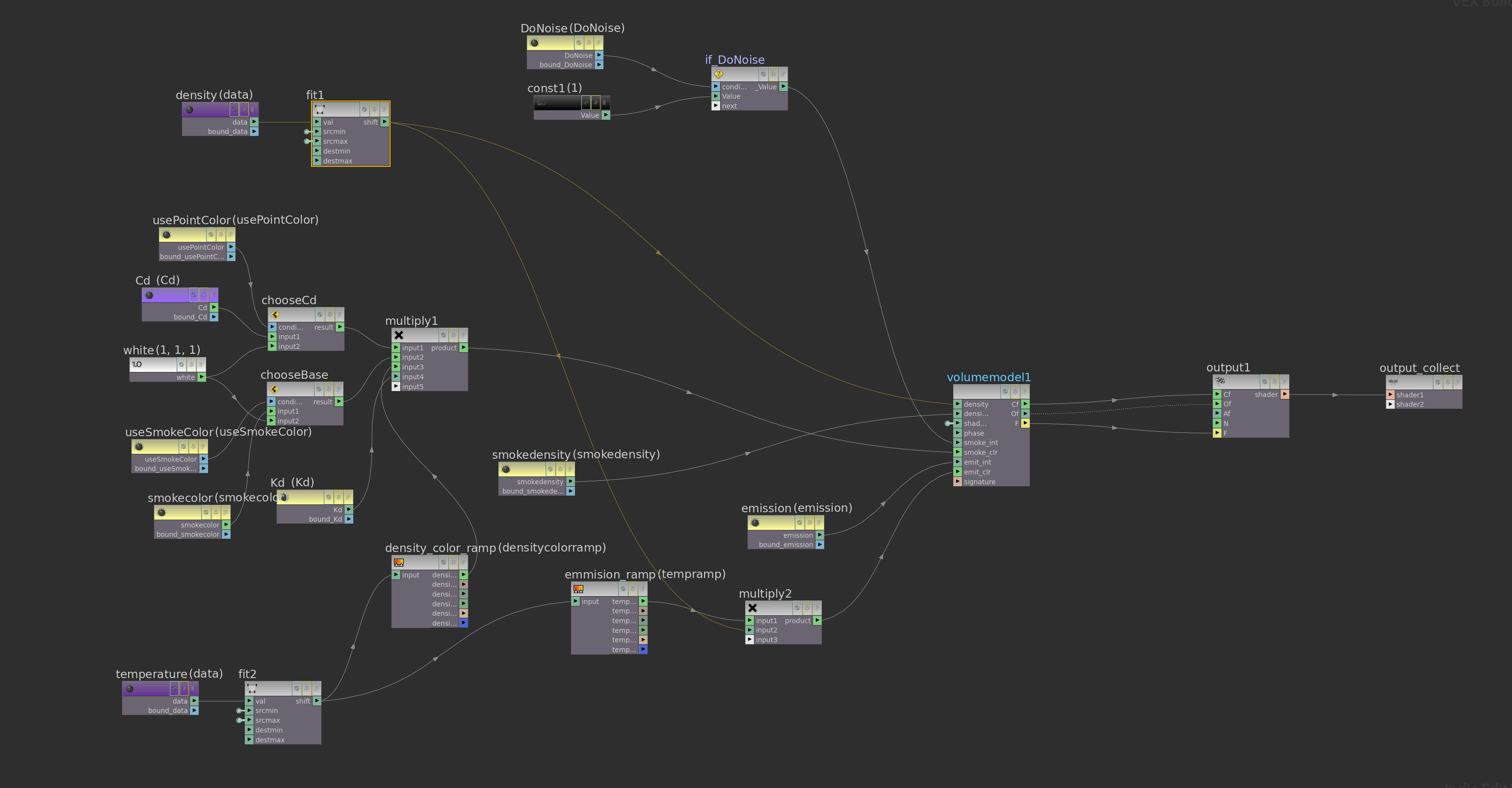}
\caption{The internal node structure of the prebuilt ``billowysmoke" shader.  Data values are shown as dark purple nodes (``density" and ``temperature") while a variety of other shader inputs and modifiers are connected together to form the basic surface colors, surface opacity and outputs of the Bidirectional Scattering Distribution Function denoted in the second most right node as Cf, Of, and F, respectively.}
\label{fig:shaderNode}
\end{figure*}

\begin{itemize}

\item {\textbf{Load Volume Data}}: {\it Houdini} accepts several graphics data formats natively, including .vdb, .geo, .bgeo, .json, .pdb, and .obj.\footnote{VDB: http://www.openvdb.org, GEO/BGEO: https://www.sidefx.com/docs/houdini11.0/io/formats/geo, JSON: http://json.org/example.html, PDB: https://github.com/Microsoft/microsoft-pdb, OBJ:  https://en.wikipedia.org/wiki/Wavefront\_.obj\_file}.  One can load in this type of data by importing through the Menu section highlighted in Figure \ref{fig:gui} or through a File node in the Network View panel, the details of which are presented in an online tutorial\footnote{www.ytini.com/tutorials/tutorial\_isolatedGalaxy.html}.  Direct access to data through the {\it yt} data reading capabilities is presented in some detail in section \ref{section:pythonsop}.

\item {\textbf{Position a Camera (Optional)}}:  One can set up test renders directly through the viewport as will be discussed in the {\textbf{Set up a Test Render}} bullet point.  However, for the sake of completeness we discuss here how one can set up a rendering camera.  The simplest way to create a static camera is by framing the scene in the Scene View window, and then creating a ``New Camera" from the ``no cam" dropdown menu at the top right corner of the Scene View. This creates a camera, looking at exactly what you see in the Scene View.   More complicated methods for positioning and pointing cameras are left to the tutorials and resources on the {\it ytini} website\footnote{www.ytini.com/tutorials/tutorial\_moreAboutCameras.html}.

\item {\textbf{Set up a Test Render}}: Once the volume and camera are created, an image can be rendered.  For a quick render, first select the ``Render View" tab above the Scene Viewer panel highlighted in Figure \ref{fig:gui}. Then select your camera (``/obj/cam1" by default), then press the {\cf Render} button.  A test render of your scene will then appear in the Scene Viewer panel in the ``Render View" tab.

\item {\textbf{Set up Shaders}}:  Shaders determine how the parameters of your dataset are translated into the opacity, color, and light reflection model passed to the image renderer.  This process can be achieved through {\it Houdini}'s CVEX\footnote{https://www.sidefx.com/docs/houdini11.1/vex/contexts/cvex} code, or through the manipulation of prebuilt SHOP (shader) nodes.  Here we outline the basic SHOP nodes to create a custom shader and leave the more complex details to a {\cf ytini.com} tutorial.  In Figure \ref{fig:shaderNode} we show an expanded view of the internal node structure of the prebuilt ``billowysmoke" shader for volume data, which is accessed through the ``Material Palette" tab in the Network View highlighted in Figure \ref{fig:gui}.  The data input parameters are shown in dark purple with the ``density" and ``temperature" labels.  These are encoded in your geometry file and imported in the {\textbf{Load Volume Data}} step of this workflow.  The user is able to modify this complex node structure with several color ramps as will be outlined in the example of section \ref{section:ajviz}.  The various other nodes control how these geometry parameters are translated into surface colors, surface opacity and outputs of the Bidirectional Scattering Distribution Function\footnote{https://www.sidefx.com/docs/houdini13.0/render/bsdf} denoted in the second most right node in Figure \ref{fig:shaderNode} as Cf, Of, and F, respectively.  As shaders can increase in complexity nearly infinitely we leave the step-by-step instructions for how to build this example shader as a tutorial\footnote{www.ytini.com/tutorials/tutorial\_moreAboutShaders.html} and provide the reader with a previously built HIP file containing this shader for direct import into {\it Houdini}\footnote{isolatedvolume\_shader.hiplc entry in table on www.ytini.com/listofHoudiniFiles.html}.

\item{\textbf{Render an Image to Disk}}: When the user has selected their final shader parameters they may render an image to a file by right-clicking in the ``Render View" tab in the Scene View panel highlighted in Figure \ref{fig:gui} and following the instructions to specify a file path and file type in which to save their image.

\end{itemize}

\section{New Methods for Importing Scientific Data To Houdini} \label{section:inputFormats}

As with the majority of production level graphics software, {\it Houdini} is not specifically designed to load observed or simulated astrophysical datasets.  Here we present two methods under development to facilitate the formatting of data from astronomy datasets to {\it Houdini} geometry nodes.  All methods described here are more fully documented on {\cf ytini.com} through tutorials and downloadable example datasets and scripts.

Both methods described in this section use {\it yt} as a means to query and format the data for use within {\it Houdini}.
The {\it yt} codebase is an open source, parallel, analysis and visualization Python package.  This widely used program includes a myriad of tools to view and analyze data including volume rendering, projections, slices, halo finding, isocontour generation and variable integration along paths.  
Here, we emphasis {\it yt}'s capabilities as a data reader and volume formatter, as it provides uniform interaction with data generated from the majority of the popular simulation codes and observational databases.
Each frontend for a specific data format in {\it yt} converts data into physically meaningful units, describes how it is stored on disk for quick and efficient reading, and prescribes a method of reading the data from disk.  
In this way, the user of {\it yt} in {\it Houdini} can remain nearly agnostic toward the specific simulation code or observational source used to generate the data.
Future plans include more direct integration of {\it yt}'s data manipulation capabilities within {\it Houdini}.   

\subsection{Direct access within Houdini through yt: A Python SOP} \label{section:pythonsop}

We begin our discussion of data access by describing an example {\it Houdini} ``SOP" which makes use of {\it yt} to read in volumetric data from a simulation, where ``SOP" is shorthand for the Surface OPerators used to construct and manipulate geometry in {\it Houdini}.  Because this Python SOP makes use of the {\it yt} libraries, one must open {\it Houdini} with their terminal's {\cf PATH} variable pointing to both {\it Houdini}'s Python libraries as well of those in {\it yt}.\footnote{Note: the Python install version of {\it Houdini} must match that of {\it yt}.  If you install with a conda or miniconda installation this can be achieved with {\cf PATH\_TO\_YT\_CONDA/bin/conda install python$=$2.7.\#} where ``\#" is the sub-version of Python installed within {\it Houdini}.  This is discussed further in the Getting Started section of the website - www.ytini.com/getstarted.html.}

\begin{figure*}
\centering
\includegraphics[width=0.8\textwidth]{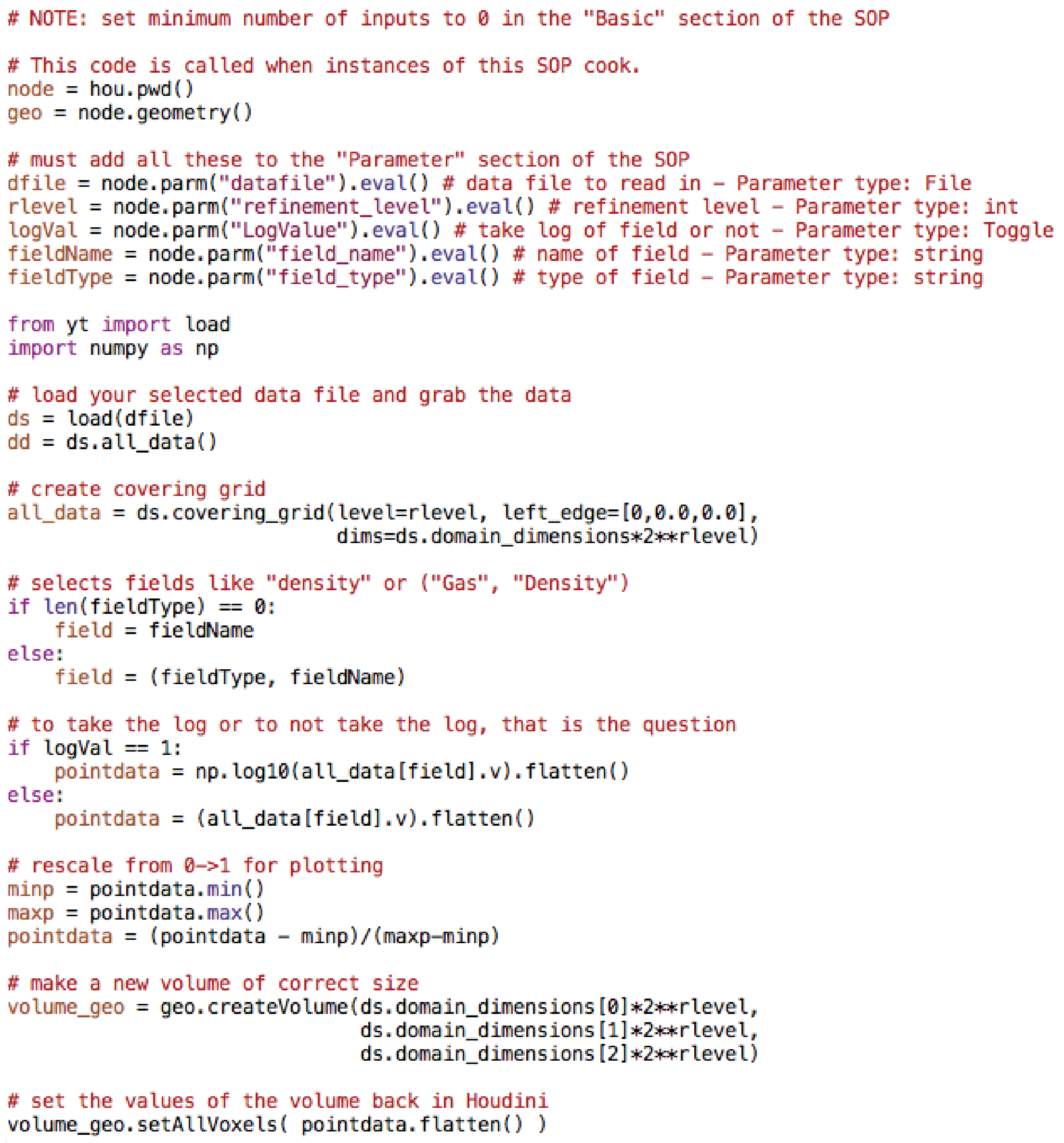}
\caption{Code for an example Python SOP loads data with {\it yt} and transforms it into a three dimensional set of voxels which are used to create a {\it Houdini} volume.  The parameters denoted with {\cf node.parm} calls allow the user to interactively change the snapshot file (datafile), refinement level of the generated covering grid (refinement\_level), whether the data should be in log values (LogValue) and the variable name to be rendered (field\_name and field\_type).}
\label{fig:pythonSOPcode}
\end{figure*}

The code for this Python SOP is shown in Figure \ref{fig:pythonSOPcode}.  This code uses several GUI-modifiable parameters to control the data file location, resolution level, and type of field to be uploaded with this SOP.  Here, {\it yt} is used as a data reader to query the simulation snapshot file and generate a uniform grid for a specific resolution, which is then formatted for {\it Houdini}'s internal Volume data format.

To add this Python SOP to one's {\it Houdini} file, one simply accesses {\cf File $\rightarrow$ New Operator Type} from the Menu section depicted in Figure \ref{fig:gui}, selects a name and label for the new Python SOP (default is ``newop" and ``New Operator", respectively), sets the {\cf Operator Style} to ``Python Type", and the {\cf Network Type} to ``Geometry Operator".  Once one clicks {\cf Accept} a new window pops up in which one can enter the various code and parameters needed for your new Python SOP.  

\begin{figure*}
\centering
\includegraphics[width=0.95\textwidth]{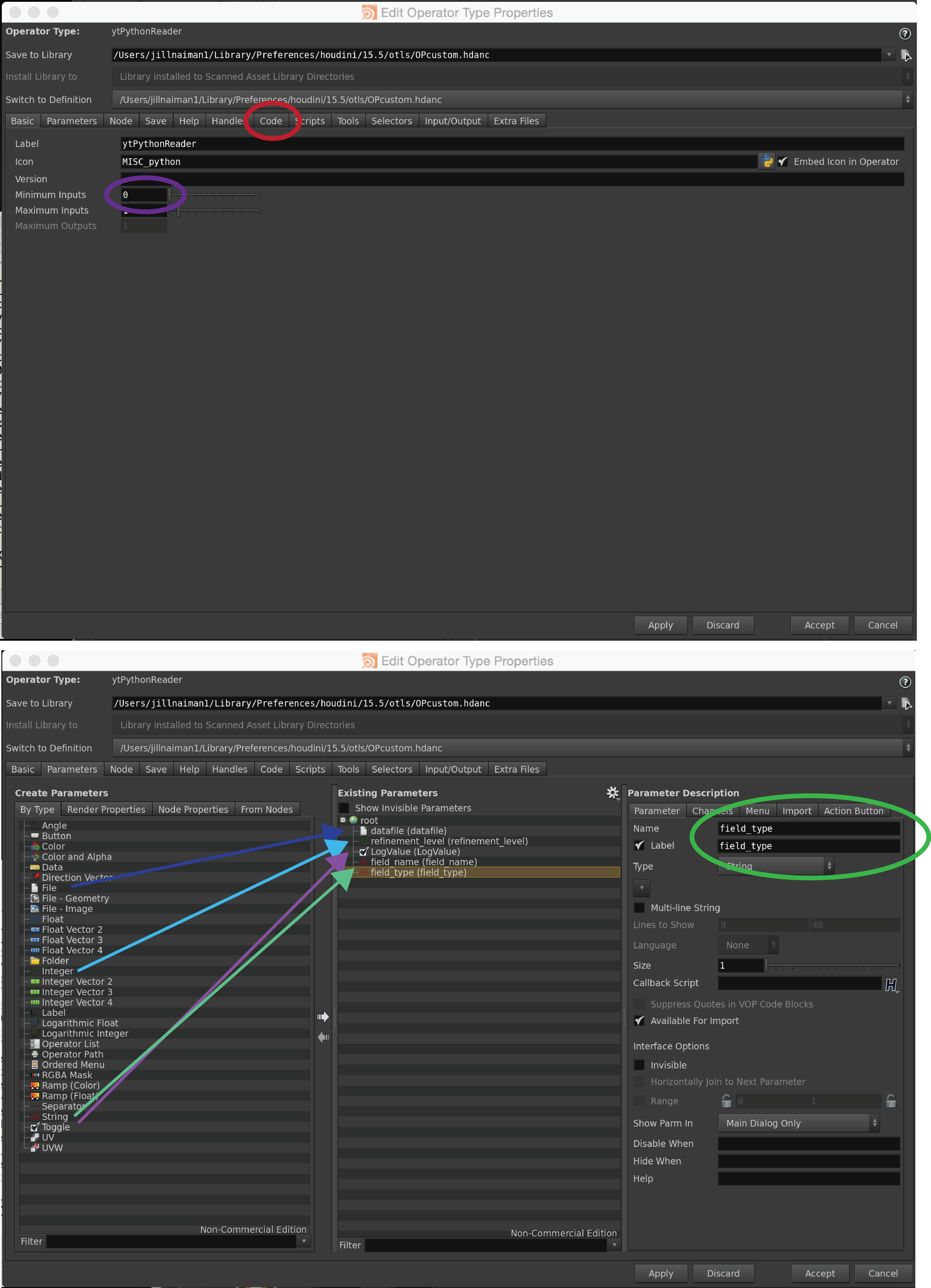}
\caption{Creating a new Python SOP first involves copying code like that shown in Figure \ref{fig:pythonSOPcode} into the code section of the Operator Type Properties pop-out window (highlighted by a red circle in the top panel).  Then the user must modify information about each one of the SOP's parameters - first by setting the minimum number of inputs to zero (highlighted by the purple circle in the top panel) and second by dragging, dropping, and naming each parameter type into the Existing Parameters list of the Parameters tab as shown with arrows and a green circle in the bottom panel.}
\label{fig:pythonSOPprocess}
\end{figure*}

The important panels necessary for this SOP import are outlined in Figure \ref{fig:pythonSOPprocess}.
First, the code from Figure \ref{fig:pythonSOPcode} must be copied into the ``Code" panel, which appears after clicking the Code panel tab highlighted by the red circle shown in the upper panel of Figure \ref{fig:pythonSOPprocess}.
In the ``Basic" panel shown in the top of Figure \ref{fig:pythonSOPprocess} one must set the default ``Minimum Inputs" to zero as no separate {\it Houdini} nodes are required to run this SOP.
In the ``Parameters" panel shown in the bottom of Figure \ref{fig:pythonSOPprocess}, one must drag-and-drop a value from the type list on the left list to the ``Existing Parameters" middle list for each parameter in the Python code outlined in Figure \ref{fig:pythonSOPcode}.  In addition, one must set the name and label of each parameter in the ``Parameter Description" section of this panel, highlighted by the green circle in Figure \ref{fig:pythonSOPprocess}.  Once the {\cf Accept} button has been selected in this New Operator Type panel, the new Python SOP is added to the current {\it Houdini} file.  

To access the SOP one simply adds a new geometry node as described in section \ref{section:workflow}, deletes the default File node within the geometry node and replaces it with one's Python SOP which is accessed within {\cf Tab $\rightarrow$ Digital Assets}.
An example HIP file with this Python SOP pre-installed is located on the file download section of {\cf ytini.com}\footnote{withPythonSOP.hipnc entry in table on www.ytini.com/listofHoudiniFiles.html}.

A more detailed tutorial with a full explanation of adding Python SOPs can be found on the ``Houdini for Astronomy" website\footnote{www.ytini.com/tutorials/tutorial\_pythonSOP.html}.

\subsection{Preprocessing Data for Houdini: The VDB Format} \label{section:vdbformat}

While the Python SOP is an excellent method for viewing smaller datasets, larger simulation files require more nuanced data handling methods.  
 The OpenVDB file format is an efficient method for storage of sparse datasets \citep{museth2013}.  Its hierarchical data structure allows OpenVDB to store high resolution, volumetric data efficiently on disk and in memory for fast I/O access.

 OpenVDB for Python includes a wrapper to access all of the VDB C++ functions within a Python interface.  Installing OpenVDB for Python and its dependencies can be accomplished through a variety of package installers (apt-get, homebrew, conda, etc).  Further details of the installation process are left to a tutorial on the ``Houdini for Astronomy" website\footnote{www.ytini.com/tutorials/tutorial\_vdbInstall.html} where a list of installation options for a variety of operating systems will be consistently updated.

\begin{figure*}
\centering
\includegraphics[width=0.8\textwidth]{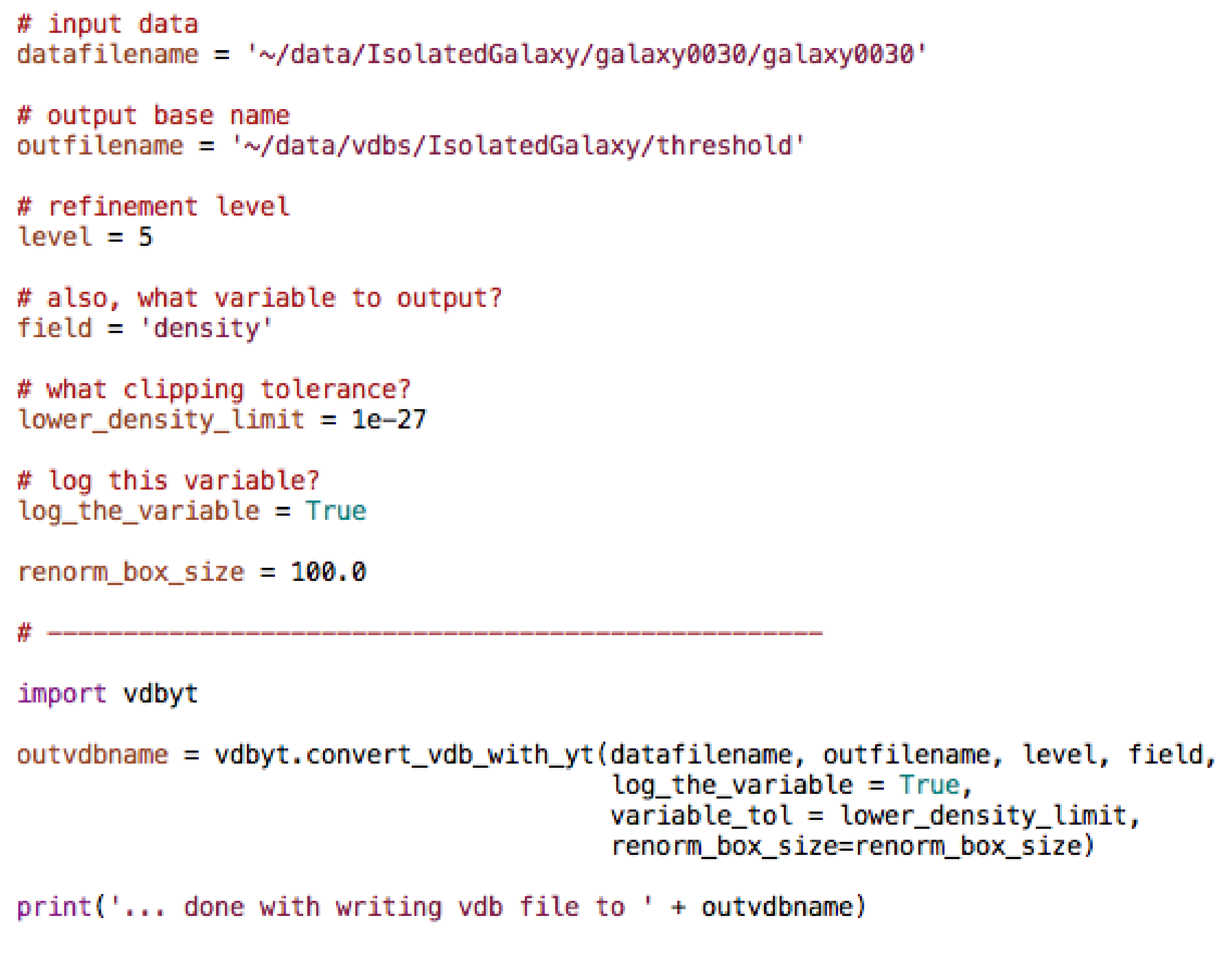}
\caption{The conversion between data of the IsolatedGalaxy simulation to VDB file format is performed with the function {\cf convert\_vdb\_with\_yt} from 
the {\cf vdbyt} Python library available on the {\it ytini} Bitbucket repository.  In this particular example, all data with a density $< 10^{-27} \, {\rm g \, cm^{-3}}$ is discarded from the output VDB vastly decreasing its occupying disk space.  The parameter {\cf renorm\_box\_size} ensures the output VDB will cover 100x100x100 {\it Houdini} units when imported.}
\label{fig:vdbconvertercode}
\end{figure*}

Figure \ref{fig:vdbconvertercode} shows an example use of the VDB converter within Python. 
Here, the function {\cf vdbyt.convert\_vdb\_with\_yt} makes use of both the {\it yt} and {\it pyopenvdb} libraries to transform any data readable with {\it yt} into the efficient VDB file format, which {\it Houdini} is optimized to read and render.
 The necessary variables include the data file, the location of the saved VDB, the refinement level of data to process, and the variable the user wishes to convert.  
 
 Several other variables are included to modify the output data.  One can choose to output the log of the variable ({\cf log\_the\_variable}, default is {\cf False}), choose to clip the output such that data with values below the clipping tolerance isn't stored in the converted VDB ({\cf variable\_tol} sets the clipping level, default is {\cf None}), choose to renormalize the minimum and maximum level of the stored data to $0 \rightarrow 1$ ({\cf renorm}, default is {\cf True}), rescale the domain size ({\cf renorm\_box}, default is {\cf True}) so it spans a specified number of {\it Houdini} domain units on import ({\cf renorm\_box\_size}, default is 10).

Figure \ref{fig:vdbconvertercode} depicts an example that employs taking the log of the output variable, disregards the storage of data with densities below $10^{-27} \, {\rm g \, cm^{-3}}$, and renormalizes the box so it 
spans 100x100x100 {\it Houdini} units upon import of the converted VDB. 
In this process, {\it yt} is utilized to read in the data and convert it to a uniform grid, sampled at the specified resolution level.  After various scaling and thresholding methods are implemented, a VDB grid is created with {\cf pyopenvdb}, and the uniform grid generated by {\it yt} is placed within the VDB grid structure.  This sparse data structure is then stored on disk at the location specified by the user.

Output of combined data sets, multi-level thresholding and further processing using {\it yt} data manipulation methods are all possible, but left for future tutorials on the ``Houdini for Astronomy" website\footnote{www.ytini.com/tutorials/tutorial\_pythonVDBconverter.html}.

\section{Example Visualizations} \label{section:examples}

As astronomical datasets grow in size and complexity the complementary methods of both visualization and analysis to inspect data are required \citep{goodman2012}.
In a similar vein to methods utilized to bring the visualization techniques of {\it Blender} to the astronomical community \citep{taylor2015,kent2015,naiman2016}, this section will focus on several methods of using {\it Houdini} to visualize astronomical data for a specific dataset from \cite{oshea2004}.  

Here, we present an outline of how one makes a production quality render from their data, checks the accuracy of their renders with one of {\it Houdini}'s data analysis nodes, and how one might combine multiple observational and computational datasets to convey a scientific concept.

\subsection{Production Quality Volume Renders} \label{section:ajviz}

 As alluded to in section \ref{section:workflow}, {\it Houdini}'s volume processing and vast shader manipulation capabilities result in the production of beautifully rendered scientific images.  
  To produce a volume render, we begin with the VDB generated from the code described in section \ref{section:vdbformat} and illustrated in Figure \ref{fig:vdbconvertercode}.  The IsolatedGalaxy dataset \citep{oshea2004} used in this example can be downloaded from the {\it yt} Data Repository website\footnote{http://yt-project.org/data/}.  The density threshold value chosen in Figure \ref{fig:vdbconvertercode} allows one to probe the in-falling gas within the galaxy's halo which surrounds an imbedded disk.
 
\begin{figure*}
\centering
\includegraphics[width=0.8\textwidth]{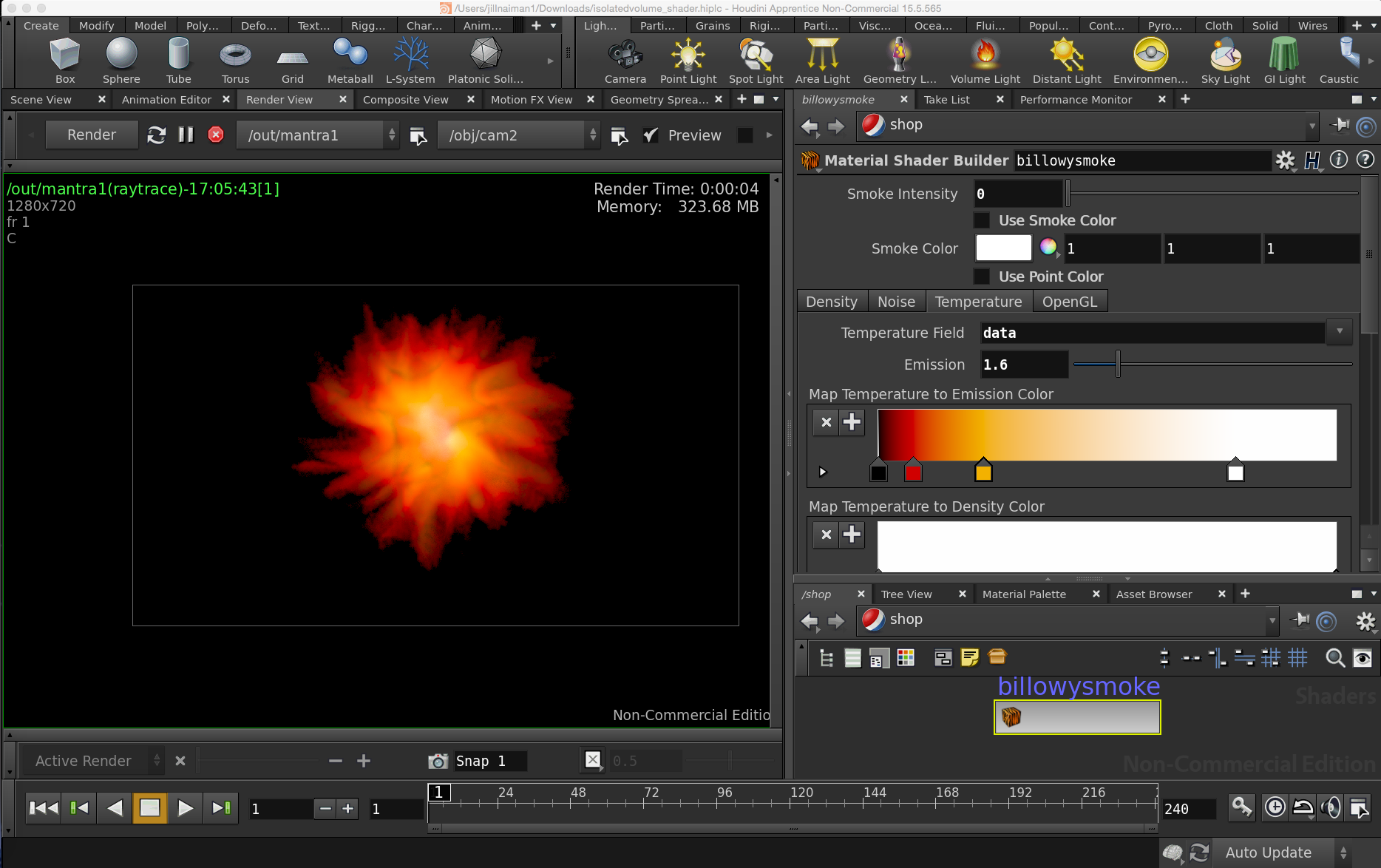}
\caption{An example of a production quality volume render from the IsolatedGalaxy dataset can be created with only minor modifications to the prebuilt shaders available in {\it Houdini}.  On the left side of this image, the render is generated using the density variable from the dataset for both opacity and emissivity color.  The right side of this image shows the various render effects and color ramps that can be used to probe and visualize different aspects of the dataset.}
\label{fig:volumerender}
\end{figure*}

  Once the VDB is imported into {\it Houdini}, we can generate the volume render of this in-falling material depicted in Figure \ref{fig:volumerender} by manipulating the shader properties of the ``billowysmoke" shader described in section \ref{section:workflow}.
 In particular, one can change the total emissivity and opacity of the entire volume, the variety of colors for each emissivity level, and a variety of ``smoke" effects to modify the render.  This allows the user to probe and highlight different features in their dataset.  
 While the image in Figure \ref{fig:volumerender} uses some defaults of the ``billowysmoke" shader to color the volume render, one can use other variables in their dataset to further modify the color scheme.

 Beyond the effects described here, this image can further be modified through {\it Houdini}'s extensive toolset for visual effects.  Complex camera paths, fading, advanced lighting, nesting of datasets, procedural noise or other added detail, and other derived features like advected field lines are all commonly used features by digital artists when using {\it Houdini} with scientific data.  Further discussion of these methods is left for future tutorials on {\cf ytini.com}.

\subsection{Accuracy Checks of Volume Renders: Analysis Plots in {\it Houdini}} \label{section:kalinaviz}

\begin{figure*}
\centering
\includegraphics[width=1.3\textwidth, angle=90]{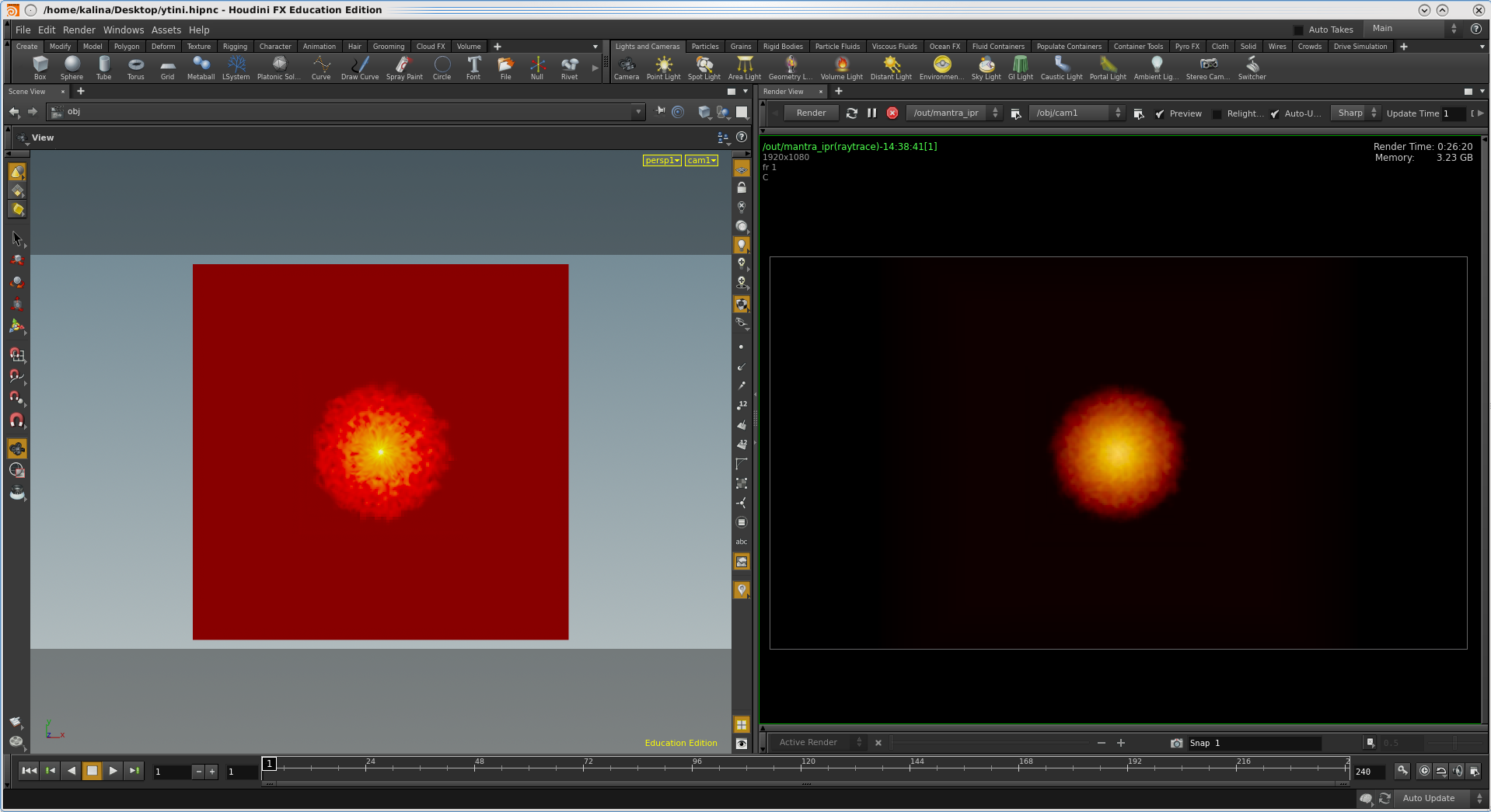}
\caption{Several different analysis plots can be utilized in {\it Houdini}, with an example of a volume slice being shown in the left panel.  One can plot different variables, slices and slice formats by adjusting the parameters in the Parameter panel.  We can compare the same dataset with a slice node in the Scene View in the left panel with its volume render using the ``billowysmoke" shader in the Render View as shown in the right panel.}
\label{fig:slice}
\end{figure*}

 While predominately a visual effects and visualization software, {\it Houdini} can also be utilized to interactively inspect datasets through interaction with the data in the Scene View and with a variety of analysis plots and tables.
 We utilize one of these features - a ``Volume Slice" node - to visualize the IsolateGalaxy dataset both using the slice plot technique many astronomers employ to analyze their simulations (Scene View in left panel of Figure \ref{fig:slice}), and the less familiar method of producing a volume render of the dataset discussed in section \ref{section:ajviz} (Render View in right panel of Figure \ref{fig:slice}).  Here, we probe gas further out from the center of the galaxy than that shown in Figure \ref{fig:volumerender} by using a lower density cut ($10^{-29} \, {\rm g \, cm^{-3}}$) in the generation of the VDB with the code depicted in Figure \ref{fig:vdbconvertercode}.

 To use this feature, the user first presses {\cf TAB} to pull up the TAB Menu in the Network View window, types {\cf volume slice}, and presses {\cf ENTER} to select. 
By connecting the output of the volume node to the input of the ``Volume Slice" node a slice across the specified volume is created as shown in the Network View of Figure \ref{fig:slice}. 
To view the slice plot instead of a render of the volume, one clicks on the center of the volume slice node to select it, followed by clicking on the right-most end of the volume slice node, to toggle the display flag. 
This will turn the right side blue indicating the volume slice will be displayed in the Scene View window.
The parameters defining the plane axis, plane offset, data attribute name, and data range can be changed in the Parameter window. 

\begin{figure*}
\centering
\includegraphics[width=0.8\textwidth]{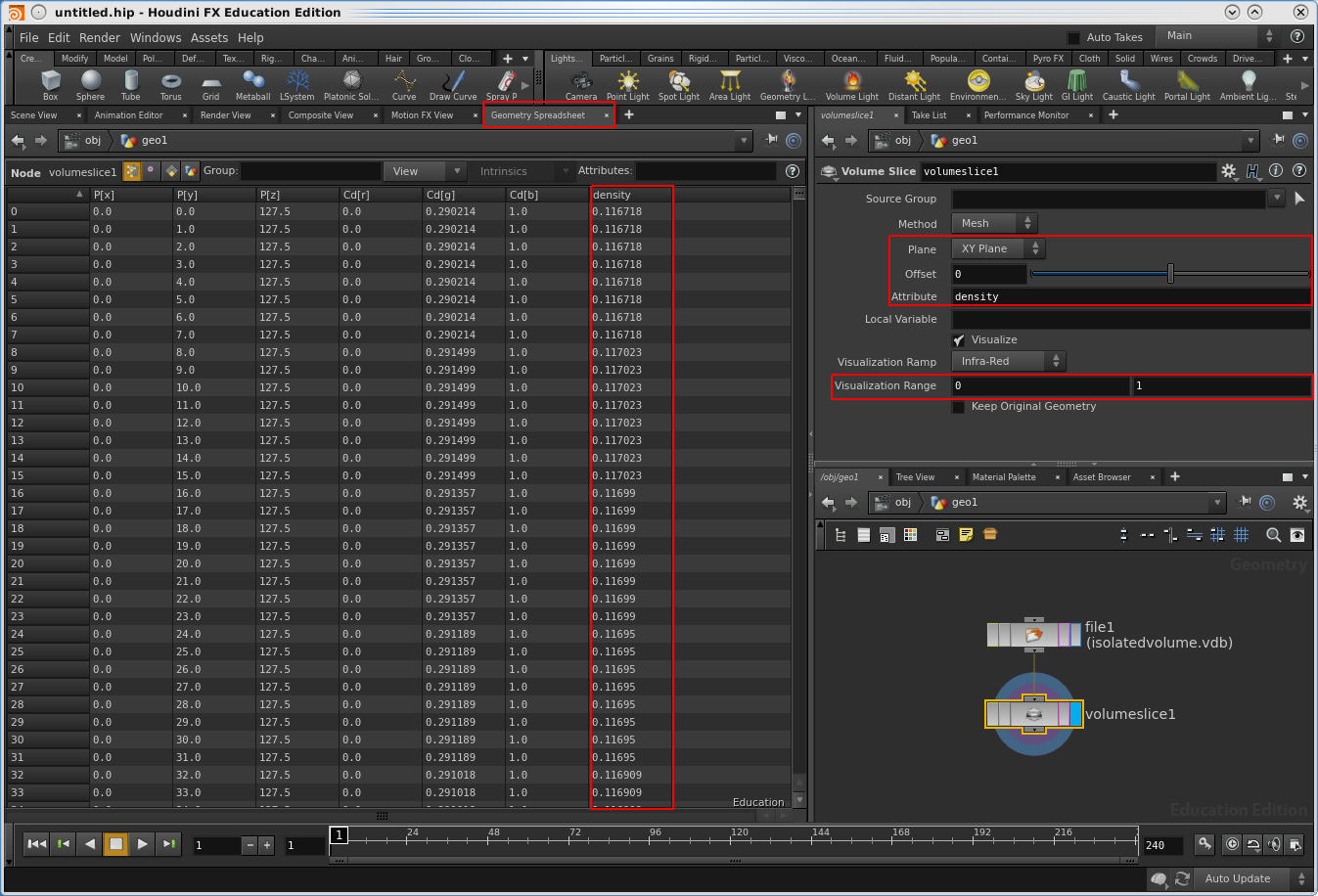}
\caption{The data along the volume slice depicted in Figure \ref{fig:slice} can be inspected further in the ``Geometry Spreadsheet" tab when a ``Volume Slice" node is connected to the geometry node as shown in the Network View.  Here, voxel number, x/y/z positions, color variables and voxel density are shown, but other parameters may be listed for a different dataset.}
\label{fig:sliceInspection}
\end{figure*}

 To inspect the individual data values along the slice, with the volume slice node selected, one switches to the Geometry Spreadsheet tab, which lives at the top of the Scene View as highlighted in Figure \ref{fig:sliceInspection}. The last column(s) of the Geometry Spreadsheet will display the data values within the slice based on their x/y/z locations (denoted by P[x], P[y] and P[z] columns). 
 In this way, the user can interact with their data in familiar (slice plot), and less familiar methods (volume render), and probe individual values of their data cube as shown in Figure \ref{fig:sliceInspection}.

 Further explanation of the analysis capabilities of {\it Houdini} will be categorized in a tutorial on {\cf ytini.com}\footnote{www.ytini.com/tutorials/tutorial\_analysisPlots.html}.  In particular, we describe using the slice plotting and volume rendering capabilities of {\it Houdini} in tandem with the IsolatedGalaxy dataset in some detail.

\subsection{Extensions: Combine and Annotate Multiple Datasets} \label{section:multiples}

\begin{figure*}
\centering
\includegraphics[width=0.8\textwidth]{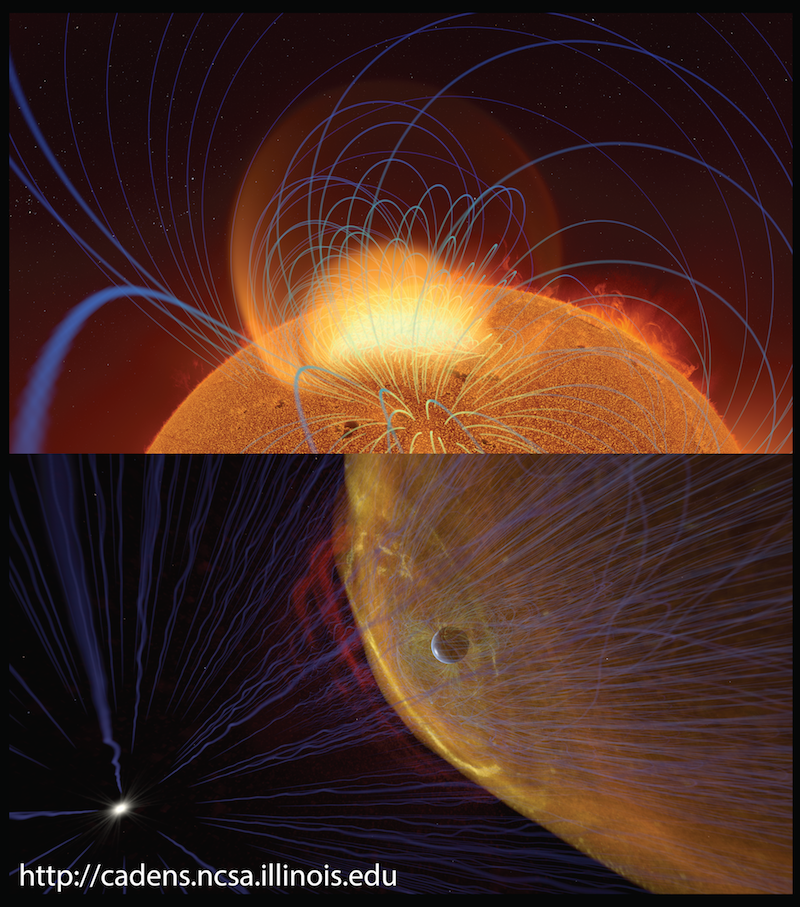}
\caption{A rendering of a coronal mass ejection (CME) as seen near the surface of the Sun is depicted in the top panel, while the CME as seen around Earth is shown in the bottom panel.  Magnetic field lines are denoted by blue and green tubes.  These images are part of the ``Solar Superstorms" documentary and dome show.}
\label{fig:suns3}
\end{figure*}

 Finally, {\it Houdini} can be utilized to combine multiple scientific datasets into dome shows and documentaries.  Figure \ref{fig:suns3} shows two frames from the full dome planetarium documentary ``Solar Superstorms"\footnote{http://www.ncsa.illinois.edu/enabling/vis/cadens/documentary/solar\_superstorms}.  
The ``Fan" image in the top panel of Figure \ref{fig:suns3} was produced from the combination of multiple data sets, as both volumes and geometrical annotations \citep{fan2016,rempel2014}. This rendering of a coronal mass ejection (CME) integrates 2D Solar Dynamics Observatory\footnote{http://sdo.gsfc.nasa.gov/} imagery around the rim of the sun; a 2D solar surface simulation, mapped onto a sphere geometry object; volumetric magnetic field lines as VDBs; and the coronal mass ejection itself as a VDB. The magnetic field line data were imported into {\it Houdini} using a Python script, and the CME was imported via a custom C++ plugin, using the {\it Houdini} Development Kit.
In the bottom panel of Figure \ref{fig:suns3} is an image from the same documentary - a rendering of solar plasma interacting with Earth's magnetic field \citep{kar2014}. The magnetic field lines were traced through the 3D simulation data using {\it yt} on the Blue Waters\footnote{http://www.ncsa.illinois.edu/enabling/bluewaters} supercomputer with a script that produced {\it Houdini}-readable .bgeo objects, which were then converted into VDB volumes. The solar plasma data were externally subsampled to reduce the file size, then read into Houdini as a .bgeo VDB object.  Full credits for these CADENS images are stored on {\cf ytini.com}\footnote{www.ytini.com/misc/SolarSuperstorms\_Magnetosphere\_Credits.pdf and www.ytini.com/misc/SolarSuperstorms\_DoubleCME\_Credits.pdf}.

 {\it Houdini}'s ability to render multiple datasets at once allows the user to compare and contrast outputs from both observational and simulated data and incorporate outputs from simulations produced with different computational methods.

\section{Summary and Future Plans} \label{section:discussion}

As the richness of observed and simulated astrophysical datasets increases with the advent of ever more powerful instruments and computers, so does the astronomer's potential to translate their scientific discoveries into visually stunning images and movies.
In this work we introduce several methods of interaction between astronomical data and {\it Houdini}, the three dimensional visual effects software used by many professional graphics studios.  
The techniques described here make use of {\it yt} as a data reader and manipulator to parse large datasets into the volume formats used by {\it Houdini}.  
We give several simple examples of processing simulation data into aesthetically pleasing images, and encourage the reader to visit the ``Houdini For Astronomy" website, {\cf www.ytini.com}, for further exploration of these methods.

Our ongoing work focuses on the volume rendering of data with non-uniform voxel sizes (e.g. generated from adaptive mesh refinement codes) which is currently beyond the capabilities of the majority of graphics and visual effects software.  Preliminary integration of AMR volume rendering is mentioned on the {\cf ytini.com} blog\footnote{http://www.ytini.com/blogs/blog\_amr\_2016-11-02.html}.  Progress on this front will be regularly updated on the website.\\
\\
The authors would like to thank Stuart Levy, Robert Patterson, Jeffrey Carpenter, Donna Cox, and Matthew Turk for illuminating conversations and Morgan Macleod and Melinda Soares-Furtado for their keen insights in regards to this paper and the anonymous referee for their extremely helpful comments.  This work is supported by a NSF grant AST-1402480, NSF award for CADENS ACI-1445176.

\bibliographystyle{apj}

\bibliography{bib_pasp_2015}

\end{document}